\documentclass[
reprint,
showpacs,
amsmath,amssymb,
prl, 
]{revtex4-1}

\usepackage{atlasphysics}
\usepackage{subfigure}
\usepackage{multirow} 
\usepackage{graphicx}
\usepackage{dcolumn}
\usepackage[hyperindex,breaklinks]{hyperref}

\usepackage[paperwidth=210mm,paperheight=297mm,centering,hmargin=1.7cm,vmargin=2.8cm]{geometry}

\usepackage{url}

\usepackage{natbib}
\newcommand\totallumieleven{$4.5~{\rm fb}^{-1}$}
\newcommand\totallumitwelve{$20.3~{\rm fb}^{-1}$}

\newcommand\htofournu{\ensuremath{H \rightarrow ZZ^{(*)} \rightarrow 4\nu}}

\newcommand\zztollnunu{\ensuremath{ZZ \rightarrow \ell\ell\nu\nu}}
\newcommand\wztolnull{\ensuremath{WZ \rightarrow \ell\nu\ell\ell}}
\newcommand\htollnunu{\ensuremath{H \rightarrow ZZ^{(*)} \rightarrow \ell\ell\nu\nu}}

\newcommand\htowwlnulnu{\ensuremath{H \rightarrow WW^{(*)} \rightarrow \ell\nu \ell\nu}}

\newcommand\mll{\ensuremath{m_{\ell\ell}}}
\newcommand\ptll{\ensuremath{p_{\mathrm{T}}^{\ell\ell}}}
\newcommand\dphill{\ensuremath{\Delta\phi(\ell,\ell)}}
\newcommand\dphizmet{\ensuremath{\Delta\phi(p_{\mathrm{T}}^{\ell\ell}, \met )}}
\newcommand\ptmiss{\ensuremath{p_{\mathrm{T}}^{\mathrm{miss}}}}

\newcommand\dphimetptmiss{\ensuremath{\Delta\phi(\met, \ptmiss)}}

\usepackage{preprintcover}  
\PreprintCoverPaperTitle{Search for Invisible Decays of a Higgs Boson Produced \\ in Association with a $\boldsymbol{Z}$ Boson in ATLAS}  
\PreprintIdNumber{CERN-PH-EP-2013-210}  
\PreprintCoverAbstract{A search for evidence of invisible-particle decay modes
of a Higgs boson produced in association with a $Z$ boson at the 
Large Hadron Collider is presented. No deviation from the Standard Model 
expectation is observed in \totallumieleven\ (\totallumitwelve) of 7~(8) TeV 
$pp$ collision data collected by the ATLAS experiment. 
Assuming the Standard Model rate for $ZH$ production, an upper limit of  
75\%, at the 95\% confidence level is set on the branching ratio
to invisible-particle decay modes 
of the Higgs boson at a mass of 125.5 GeV. The limit 
on the branching ratio is also interpreted in terms of an upper limit on the 
allowed dark matter--nucleon scattering cross section within a Higgs-portal 
dark matter scenario. 
Within the constraints of such a scenario, the results presented in this Letter 
provide the strongest available limits for low-mass dark matter candidates.  
Limits are also set on an additional neutral Higgs boson, in the mass 
range $110<m_H<400$~GeV, produced in association with a $Z$ boson and decaying 
to invisible particles.}  
\PreprintJournalName{Phys. Rev. Lett.}  

\begin{document}

\title{Search for Invisible Decays of a Higgs Boson Produced \\ in Association with a $\boldsymbol{Z}$ Boson in ATLAS}

\author{G. Aad \it{et al.}}
\collaboration{ATLAS Collaboration}

\date{\today}


\begin{abstract}

A search for evidence of invisible-particle decay modes 
of a Higgs boson produced in association with a $Z$ boson
at the Large Hadron Collider is presented. 
No deviation from the Standard Model expectation is observed 
in \totallumieleven\ (\totallumitwelve) of 7~(8) TeV 
$pp$ collision data 
collected by the ATLAS experiment. 
Assuming the Standard Model rate for $ZH$ production, an upper limit of  
75\%, at the 95\% confidence level is set on the branching ratio
to invisible-particle decay modes 
of the Higgs boson at a mass of 125.5 GeV. The limit 
on the branching ratio is also interpreted in terms of an upper limit on the 
allowed dark matter--nucleon scattering cross section within a Higgs-portal 
dark matter scenario. 
Within the constraints of such a scenario, the results presented in 
this Letter provide the strongest available limits for low-mass 
dark matter candidates.
Limits are also set on an additional neutral Higgs boson, in the mass 
range $110<m_H<400$~GeV, produced in association with a $Z$ boson and decaying 
to invisible particles.

\end{abstract}

\maketitle

Some extensions of 
the Standard Model (SM) allow a 
Higgs boson~\cite{Englert:1964et,Higgs:1964pj,Guralnik:1964eu}~to 
decay to a pair of stable or long-lived particles~\cite{Hinv:Griest1988,Hinv:Djouadi1996,Hinv:Antoniadis2005,Hinv:Binoth1997,Hinv:Chikashige1981,Hinv:Joshipura1992,Hinv:Shrock1982,Hinv:Mohapatra1986,Hinv:Gonzalez1989,Hinv:Carlson1989,Hinv:Li1985,Hinv:Zee1980,Hinv:Campos1997,Hinv:Martin1999,Hinv:Giudice2001}
that are not observed by the ATLAS detector. 
For instance the Higgs boson can decay into two particles with very 
small interaction cross sections with SM particles, such as dark matter (DM) candidates. 
Collider data can be used to directly constrain the branching ratio of 
the Higgs boson to invisible particles.
Similarly, limits can be placed on the 
cross section times branching ratio 
of any additional Higgs bosons
decaying predominantly to invisible particles. 
LEP results~\cite{LEP} put 
limits on an invisibly decaying Higgs boson, produced in association with 
a $Z$ boson, for Higgs masses below 120~GeV.

This Letter presents a search for invisible decays of 
a Higgs boson produced in association with a $Z$ boson, 
as suggested in Refs.~\cite{Hinv:Choudhury1994,Hinv:Godbole2003,Hinv:Davoudiasl2005}.
A Higgs boson in the mass range $110<m_H<400$~GeV is considered. 
The distribution of the missing transverse momentum (\met) in events with an electron 
or a muon pair consistent with a $Z$ boson decay
is used to constrain the $ZH$ production cross section times the branching ratio of the 
Higgs boson decaying to invisible particles, over the full mass range.  
For the newly discovered Higgs boson, 
a constraint could be placed on the branching ratio to invisible particles. 
In this case the mass of the Higgs boson is taken to be $m_H=125.5~\gev$, 
the best-fit value from the ATLAS experiment~\cite{Aad:2013wqa}, 
and the $ZH$ production cross section is assumed to be that predicted for the SM Higgs boson. 
This assumption 
implies that the 
hypothesized unobserved particles that couple to the Higgs boson have sufficiently 
weak couplings to other SM particles 
to not affect the Higgs boson production cross sections.   
The total cross section for the associated production of a SM Higgs boson, with $m_H=125.5$~GeV, 
and a $Z$ boson, calculated to next-to-next-to-leading order in QCD~\cite{Brein:2003wg} 
and including next-to-leading-order (NLO) electroweak corrections~\cite{Ciccolini:2003jy,Denner:2011id}, 
is 331~fb at $\sqrt{s} = 7$~TeV and 410~fb at $\sqrt{s} = 8$~TeV~\cite{LHCHiggsYR3}. 
The SM branching ratio of the Higgs boson decaying to invisible particles 
is $1.2\times 10^{-3}$, arising from the \htofournu\ decay.
The present search is not sensitive to the low branching ratio for this decay, 
but instead searches for enhancements in the decay fraction to invisible particles due
to physics beyond the Standard Model (BSM).

The search uses \totallumieleven\ of data recorded with the ATLAS detector in 2011 at  
$\sqrt{s} = 7$~TeV and \totallumitwelve\ of data recorded in 2012 at  $\sqrt{s} = 8$~TeV.  
The ATLAS detector has been described elsewhere~\cite{Aad:2008zzm}.
Simulated signal and background event samples are produced with Monte Carlo (MC) event 
generators, passed through a full GEANT4~\cite{geant4} simulation of the ATLAS 
detector~\cite{simuAtlas} and reconstructed with the same software as the data.

The signal samples are generated with Herwig++~\cite{herwigpp}
and its internal POWHEG method~\cite{powheg1,powheg2}.
The SM $ZZ$ and $WZ$ backgrounds are taken from simulation, since they 
have limited statistics in the control regions that would allow to estimate these
backgrounds with data. 
All the other background processes to this search are determined from data. 
In these cases, 
simulated samples are only used as cross-checks for the obtained background estimates. 
POWHEG~\cite{powheg1,powheg2,powheg3} interfaced with PYTHIA8~\cite{Sjostrand:2007gs} is 
used to model SM $ZZ$ and $WZ$ production~\cite{powheg_ew}. The production of $WW$ 
is modeled using HERWIG~\cite{herwig} and SHERPA~\cite{sherpa} for the 7 and 8 TeV data, 
respectively. A separate sample simulated with gg2VV~\cite{gg2ww}  interfaced with 
JIMMY~\cite{jimmy} accounts for $WW/ZZ$ 
production through quark-box diagrams, which are not included in the above mentioned samples.  
The MC@NLO~\cite{mcatnlo} generator interfaced with JIMMY is used to model 
$t\bar{t}$, $Wt$, and {\it s}-channel single top-quark production.  
AcerMC~\cite{acermc} interfaced with PYTHIA~\cite{pythia} models {\it t}-channel single 
top-quark production.
Inclusive $Z/\gamma$* production is simulated with ALPGEN~\cite{alpgen} interfaced with 
JIMMY or PYTHIA for the 7 or 8~TeV data, respectively.  
Inclusive $W$ production is simulated with ALPGEN interfaced with JIMMY. 
Contributions to this search from the 
\htowwlnulnu~and \htollnunu~decays 
of a 125.5~GeV SM Higgs boson are studied
using POWHEG~\cite{powheg1,powheg2,powheg3,powheg4,powheg5} interfaced with PYTHIA8 and 
found to be negligible.

Electron candidates are reconstructed from isolated energy deposits in the electromagnetic calorimeter with a 
shower shape consistent with electrons or photons, matched to inner detector tracks~\cite{Aad:2011mk}.  
The electrons used to form a $Z$ boson candidate are required to have transverse momentum $\pt > 20 \gev$ 
and pseudorapidity $|\eta| < 2.47$~\cite{AtlasCoordinateSystem}. 
Electrons with $\pt > 7 \gev$ that satisfy less stringent identification criteria on the 
calorimeter cluster shape, track quality, and track-cluster matching~\cite{Aad:2011mk} are used to veto 
events with more than two charged leptons.  

Muon candidates are reconstructed combining tracks independently found in the muon spectrometer and 
inner tracking detector~\cite{Aad:2010yt}.  
Muons forming  a $Z$ boson candidate are required to have $\pt > 20 \gev$ and $|\eta| < 2.5$.  
Muons with $\pt > 7 \gev$ 
are used to veto events with more than two charged leptons.

Jets are reconstructed using the anti-$k_{t}$ algorithm~\cite{antikt} with a radius parameter $R = 0.4$.
They must have $\pt > 20 \gev$ and $|\eta| <$ 4.5.  To discriminate against jets from 
additional minimum bias interactions,
selection criteria are applied to ensure that most of the jet momentum, 
for jets with $|\eta| < 2.5$, is associated with tracks originating from  
the primary vertex, which is taken to be the vertex with the highest 
summed $p_{\mathrm{T}}^2$ of associated tracks. 

To ensure good separation between electrons, muons and jets, 
electrons are removed if they are within $\Delta R \leq 0.2$ of an 
identified muon, and jets are removed if they are within $\Delta R \leq 0.2$ 
of an identified electron. Remaining electrons and muons are removed if 
they are within $\Delta R \leq 0.4$ of a remaining jet or if the scalar 
sum of track momenta, not associated with the lepton, in a cone of 
$\Delta R < 0.2$ around the 
lepton direction is greater than 10\% of the lepton \pt.

The \met~is the magnitude of 
the negative vectorial sum of the transverse momenta from calibrated objects, 
such as identified electrons, muons, photons, hadronic decays of tau leptons, and jets~\cite{metcalc}.  
Clusters of calorimeter cells not matched to any object are also included. 
The analysis also uses a track-based missing transverse momentum 
(\ptmiss) 
computed from all 
inner detector tracks with $\pt > 500 \mev$ and $|\eta|<2.5$, 
that satisfy stringent quality 
criteria~\cite{trackmetCONF} and are consistent with originating from the primary vertex. 
For the \ptmiss~calculation, tracks matched to electrons are discarded 
and replaced by the transverse energy \ET\ of the matched cluster 
measured in the calorimeter to include any photon radiation in the 
calculation.

Event selection criteria are determined in an optimization procedure, 
using simulated samples, to maximize the signal significance of the search. 
Events are required to pass a single-lepton or lepton-pair trigger, with 
small variations in the applied \pt\ threshold in different data-taking periods.
Events must also have at least one 
reconstructed vertex with at least three associated tracks with 
$\pt > 500~\mev$.  
Data quality criteria are applied to reject events from non-collision 
backgrounds 
or events with degraded detector performance~\cite{metcalc}. 

The invariant mass of the selected dilepton system, \mll, is required to 
satisfy $76<\mll<106 \gev$ to be consistent with leptons originating from a $Z$ boson decay.
\begin{figure}[t]
  \begin{center}
    \includegraphics[width=0.49\textwidth]{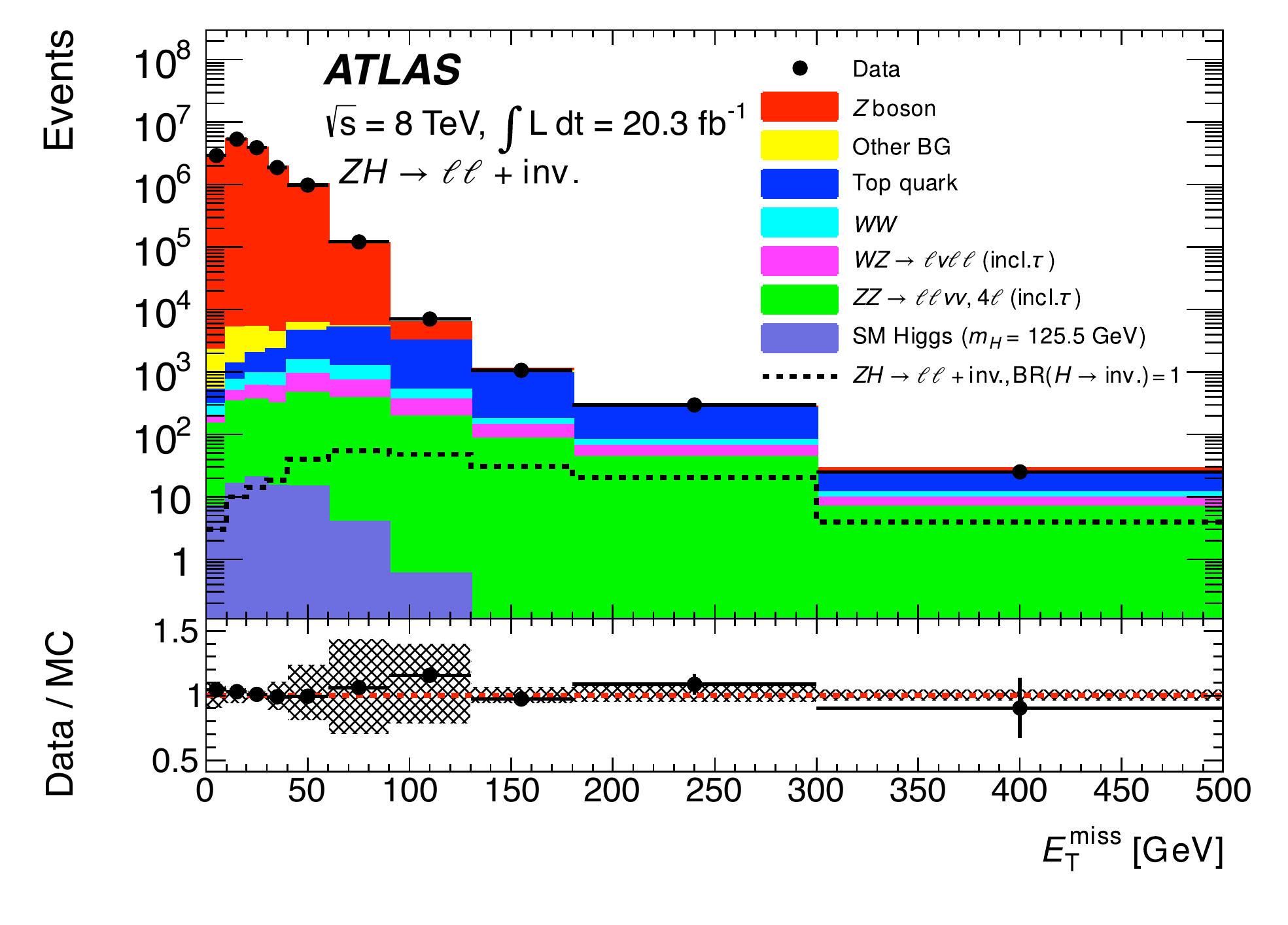}
  \end{center}
  \caption{\label{fig:met_after_Mll} Distribution of \met~for events with 
  the invariant mass of the two leptons $76<\mll<106 \gev$
  in the 8 TeV data (dots). The stacked histograms represent the background
  predictions from simulation. The signal hypothesis is shown by a dotted line and assumes 
  the SM $ZH$ production rate for a $m_H = 125.5 \gev$ Higgs boson 
  with $\text{BR}(H\to \text{inv.})=1$. The inset at the bottom of the figure shows the ratio of 
the data to the combined background expectations as well as a band corresponding to the combined systematic uncertainties.}
\end{figure}

Figure~\ref{fig:met_after_Mll} shows the \met\ distribution in the 8 TeV data 
sample after the dilepton mass requirement. 
In this figure the data are consistent with the expected background based on simulated samples for all 
but the multijet background. 
The uncertainty band of the expected background is widest in the region dominated by the steeply 
falling $Z$ boson background.
To reject the majority of this background, \met\ is
required to be greater than 90 GeV.  
In events where a significant \met\ arises from mis-reconstructed energy in the calorimeter,
the vectors of \met~and \ptmiss~are 
likely to have different azimuthal angles.
Thus the azimuthal difference of these
two vectors, \dphimetptmiss, is required to be less than 0.2. 

For the signal, the momentum of the reconstructed $Z$ boson is expected to be balanced 
by the momentum of the invisibly decaying Higgs boson. Therefore
the azimuthal separation between the dilepton system 
, where the magnitude of its transverse momentum is defined as \ptll,
and the \met, \dphizmet,
is required to be greater than 2.6.  
The boost of the $Z$ boson causes the decay leptons to be produced with a small opening angle. The 
azimuthal opening angle of the two leptons, \dphill, is thus required to be less than
1.7. 
Furthermore 
\ptll\ and \met\ are expected to be similar. Therefore 
the fractional \pt\ difference, defined as $|\met - \ptll|/\ptll$, is required to be
less than 0.2.   
Finally, for the majority of the signal no additional high-\pt\ jets are expected to be observed in 
the events, while for the background from boosted $Z$ bosons 
and from $t\bar{t}$ pairs one or more jets are expected. Thus, events are required to have 
no reconstructed jets with $\pt > 25 \gev$ and $|\eta|<2.5$.

\begin{table*}[t] 
  \centering
  \begin{tabular}{lccc} 
    \hline\hline
     Data Period    & 2011 (7 TeV) & 2012 (8 TeV) \\
    \hline
      \zztollnunu                            &  $20.0 \pm 0.7 \pm 1.6$    & $91 \pm 1 \pm 7$  \\ 
      \wztolnull                               &  $4.8 \pm 0.3 \pm 0.5$     & $26 \pm 1 \pm 3$  \\ 
      Dileptonic $t\bar{t}$, $Wt$, $WW$, $Z\to\tau\tau$ &  $0.5 \pm 0.4 \pm 0.1$ & $20 \pm 3 \pm 5$  \\ 
      $Z\to ee$, $Z\to\mu\mu$               &  $0.13 \pm 0.12 \pm 0.07$  & $0.9 \pm 0.3 \pm 0.5$  \\ 
      $W$ + jets, multijet, semileptonic top  &  $0.020 \pm 0.005 \pm 0.008 $     & $0.29 \pm 0.02 \pm 0.06$  \\ 
      Total background                            &  $25.4 \pm 0.8 \pm 1.7$     & $138 \pm 4 \pm 9$  \\ 
      \hline
      Signal ($m_H=125.5$ GeV, $\sigma_{ZH,\mathrm{SM}}$, $\text{BR}(H\to \text{inv.})=1$) & $8.9 \pm 0.1 \pm 0.5$ & $44 \pm 1 \pm 3$ \\
      \hline
      Observed        & 28 & 152  \\
      \hline\hline
  \end{tabular}
  \caption{\label{tab:eventstable_llinv} Number of events observed in data and expected from the signal and from each background source for the 7 and 8 TeV data-taking periods. Uncertainties on the signal and background expectations are presented with statistical uncertainties first and systematic uncertainties second.}
\end{table*}

After the selection requirements, the dominant background is SM $ZZ$ production
followed by SM $WZ$ production, 
as shown in Table~\ref{tab:eventstable_llinv}.
These 
backgrounds are simulated using MC samples normalized to NLO cross sections.
The simulation of $WZ$ events is validated by comparing them to data events in which 
the third-lepton veto is replaced by an explicit third-lepton requirement.
The theoretical prediction of the $ZZ$ production is in agreement 
with the ATLAS 
cross-section measurement at $\sqrt{s} = 7$~TeV~\cite{Aad:2013zz}. 

Background contributions from 
events with a genuine isolated lepton pair, not originating from a 
$Z\rightarrow ee$ or $Z\rightarrow \mu\mu$ decay ($WW$, $t\bar{t}$, $Wt$, and $Z\to\tau\tau$), 
are estimated
by exploiting the flavor symmetry in the dilepton final state of these processes.
Distributions for events with an $e\mu$ pair, appropriately scaled to account for   
differences in electron and muon reconstruction efficiencies, can be used to estimate 
this background in the electron and muon channels. 
The difference between the 
efficiencies for electrons and muons is estimated using the square root of 
the ratio of the numbers of dimuon and dielectron events in 
data within the \mll\ window.
Events in the $e\mu$ control region not originating from 
$WW$, $t\bar{t}$, $Wt$, or $Z\to\tau\tau$ backgrounds 
are subtracted using simulated samples.
Important sources of systematic uncertainty are variations in the 
correction factor for the 
efficiencies for electrons and muons 
and uncertainties in the simulated samples used for the subtraction.
The combined systematic uncertainty is 23\% for both the 7 and 8 TeV data. 
The estimated background from these sources is consistent with 
the expectation from the simulation.

The background from inclusive 
$Z\rightarrow ee$ and $Z\rightarrow \mu\mu$ 
production  
in the signal region is estimated from the background in three sideband regions~\cite{Aad:2010ey}.
These sideband regions are formed by considering events failing one or both of the
nominal selection requirements applied to \dphimetptmiss\ and the fractional \pt\ difference. 
Contributions from non-$Z$ backgrounds in the sideband regions are subtracted.
The impact from a correlation between the above two variables is determined from the simulation and 
a correction, of at most 7\%, is applied to account for it.
The main uncertainties are due to variations in this correction and differences in the shape 
of the \met\ distribution in the control regions. The overall systematic uncertainty is 
52\% in the 7 TeV data and 59\% in the 8 TeV data. 

The small background from events with only one genuine isolated lepton (inclusive $W$, single-lepton top 
pairs and single top production) or from multijet events 
are estimated from data using control samples, selected by requiring two lepton candidates of which 
at least one fails the full lepton selection criteria. These samples are scaled with
a measured \pt-dependent factor, determined from data as described in Ref.~\cite{Aad:2011kk}.
Systematic uncertainties are determined following the procedures used in Ref.~\cite{Aad:2011kk},
yielding an uncertainty of 40\% in the 7 TeV data 
and 21\% in the 8 TeV data.

Systematic uncertainties on the signal and the SM $ZZ$ and $WZ$ backgrounds are derived from the 
luminosity uncertainty, the propagation of
reconstructed object uncertainties and from theoretical uncertainties on the production
cross sections. 
The luminosity uncertainty is 1.8\% 
for the 7 TeV data-taking period and 2.8\% for the 8 TeV data-taking period~\cite{Aad:2013ucp}. 

Lepton trigger and identification efficiencies as well as the energy scale and resolution are 
determined from data using large samples of 
$Z$ events. After appropriate corrections to the simulation, 
uncertainties are propagated to the event selection. 
These uncertainties contribute typically 1.0--1.5\% to the overall 
selection uncertainty. 
Jet energy scale and resolution uncertainties are derived 
using a combination of techniques that use dijet, 
photon + jet, and $Z$ + jet events~\cite{jes11,jer11}. 
These contribute an uncertainty of between 3\% and 6\% on the final 
event selection.
The uncertainties on the energy scale and resolution of leptons and jets are also propagated 
to the \met\ calculation, 
and the resulting 
uncertainty in the latter is included in uncertainties given above.
Uncertainties in the pile-up simulation, affecting in particular \met, contribute a 
further 1--2\% uncertainty.

Theoretical uncertainties on the $ZH$ production cross section are derived from variations of 
the renormalization and factorization scale, $\alpha_{s}$, and the parton distribution functions (PDFs)
~\cite{LHCHiggsYR3}.
These are combined to give an uncertainty of 3.6--5.7\% on the cross section. 
This analysis is sensitive to the distribution of the Higgs boson \pt\ through the \met, and 
uncertainties in the \pt\ boost of the Higgs boson can affect the signal yield. 
An additional systematic uncertainty of 1.9\% is applied to the 
normalization~\cite{Ciccolini:2003jy,Denner:2011id,atlasHbb}, 
and uncertainties as a function of the Higgs boson \pt\ are considered as a systematic shape uncertainty.

The cross-section uncertainty on the 
$ZZ$ background is 5\% from varying the PDFs, $\alpha_{s}$, and QCD scale. 
The uncertainty on the jet veto for the $ZZ$ background 
due to the parton showering is estimated 
to be  6.4\% (5.5\%) for the 7 (8) TeV data.
Because the \met~distribution of the final selected sample is used in the 
limit-setting procedure, 
the impact of PDFs, $\alpha_{s}$, and QCD scale uncertainties on the shape of 
this distribution is also considered. 
The theoretical uncertainty of the $WZ$ background is considered similarly. 
The total systematic uncertainty on the SM $ZZ$ background is 
8\% for both the 7 and 8 TeV data-taking periods, whereas for the 
$WZ$ background it is 10\% (13\%) for the 7 (8) TeV data-taking periods.

Event reconstruction and theoretical uncertainties are considered as 
correlated between the 7 and 8 TeV 
data, and between the signals and backgrounds estimated from simulation. 
The systematic uncertainties in 
methods that determine backgrounds from data using control regions
are also assumed to be correlated between the two datasets.  
The luminosity uncertainty is considered as uncorrelated between the 7 and 8 TeV data.

\begin{figure}[t]
  \begin{center}
    \includegraphics[width=0.49\textwidth]{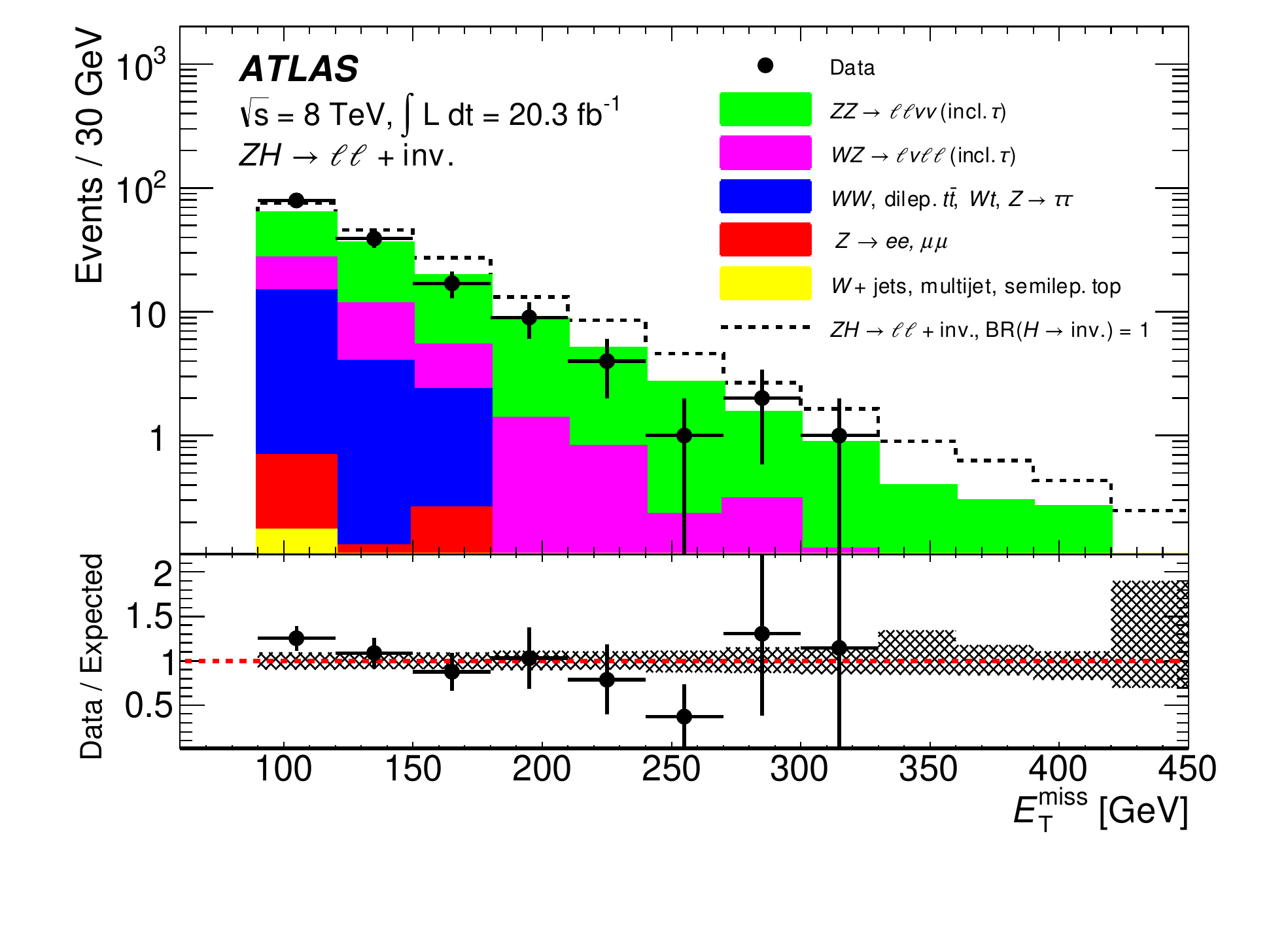}
  \end{center}
  \caption{\label{fig:final_met_125} Distribution of \met\ after the full selection in the 8 TeV data (dots).  
                                        The filled stacked histograms represent the background expectations. 
                                        The signal expectation for a Higgs boson with $m_H=125.5$~GeV,
                                        a SM $ZH$ production rate and  $\text{BR}(H\to \text{inv.})=1$ is
					stacked on top of the background expectations. 
                                        The inset at the bottom
                                        of the figure shows the ratio of the data to the combined
                                        background expectations. The hashed area shows the systematic 
                                        uncertainty on the combined background expectation. 
                                           }
\end{figure}

The numbers of observed and expected events for the 7 and 8 TeV data-taking
periods are shown in Table~\ref{tab:eventstable_llinv}. 
Figure~\ref{fig:final_met_125}
shows the \met\ distribution after the full event selection for the 8 TeV data and the expected
backgrounds. 
The normalization of the backgrounds is extracted from a binned profile maximum likelihood fit
in the signal region. Systematic uncertainties are considered as nuisance parameters, and 
are assumed to be constrained by Gaussian distributions. 
The signal expectation shown corresponds to a Higgs boson with $m_H=125.5$~GeV, a SM $ZH$ production rate 
and $\text{BR}(H\to \text{inv.})=1$.
No significant excess is observed over the SM expectation.

\begin{figure}[t] 
  \begin{center}
   \includegraphics[width=0.49\textwidth]{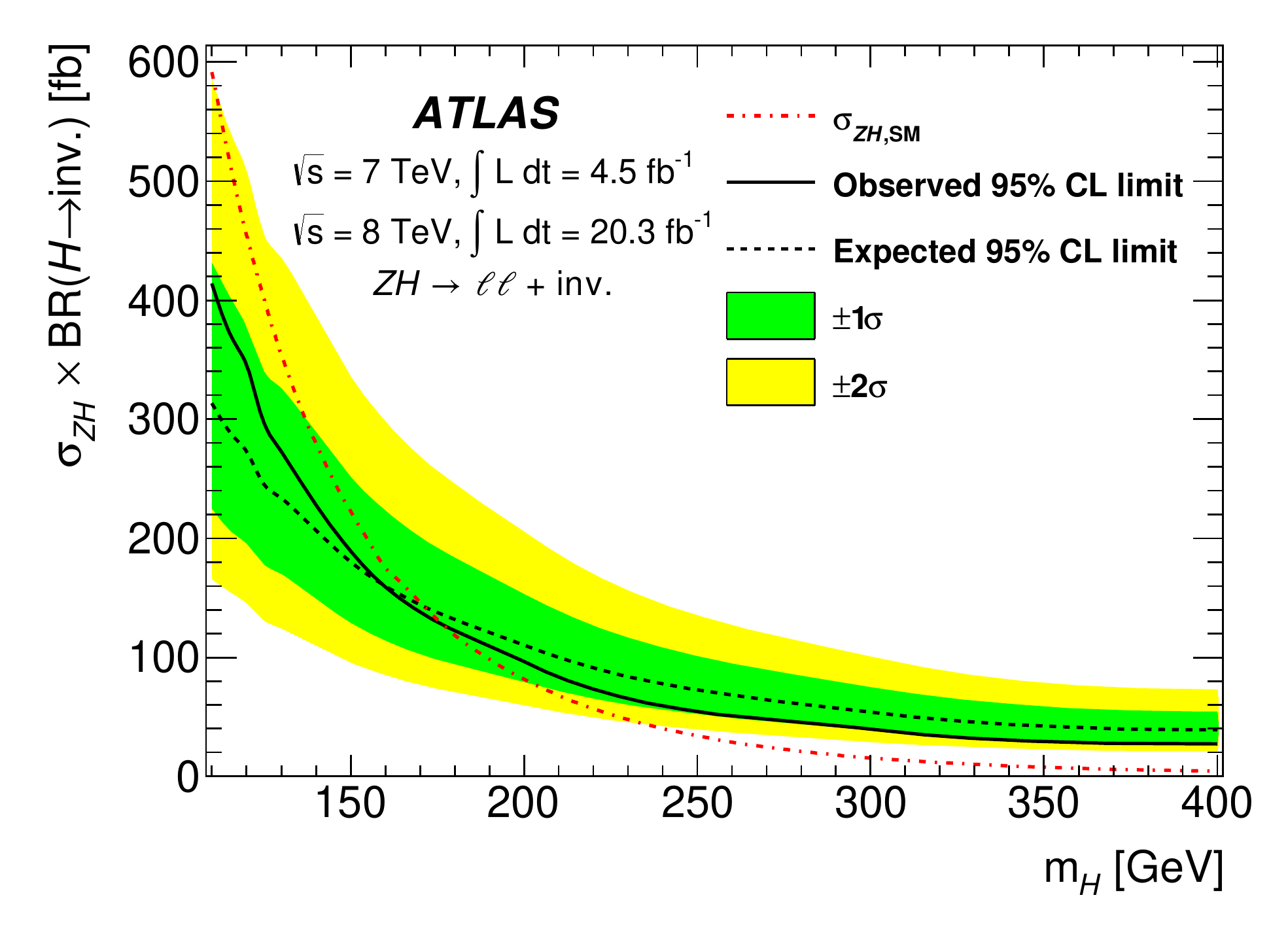}
  \end{center}
\caption{\label{fig:limits} Upper limits on $\sigma_{ZH}\times \text{BR}(H\to \text{inv.})$ at 95\% CL for a Higgs boson 
with $110<m_H<400$~GeV, for the combined 7 and 8 TeV data. The full and dashed lines show the observed and expected limits, respectively.}
\end{figure}

Limits are set on the cross section times branching ratio 
for a Higgs boson decaying to invisible particles anywhere in the mass range 
$110<m_H<400 \gev$. 
The limits are computed using a maximum likelihood fit to the \met~distribution
following the $CL_s$ (signal confidence level) modified frequentist formalism~\cite{CLspaper} with a
profile likelihood test statistic~\cite{asymptotic}. 
Figure~\ref{fig:limits} shows the 95\% CL upper limits on $\sigma_{ZH}\times \text{BR}(H\to \text{inv.})$ in the mass range 
$110<m_H<400$~GeV for the combined 7 and 8 TeV data. The expectation for a Higgs boson with a production cross section 
equal to that expected for a SM Higgs boson
and $\text{BR}(H\to \text{inv.})=1$ is also shown.

For the discovered Higgs boson an upper limit of 75\% at 95\% CL (63\% at 90\% CL) is set on the
branching ratio to invisible particles. For this the predicted SM $ZH$ production rate 
with $m_H$ = 125.5 GeV, is assumed. The expected limit in the absence of BSM decays to invisible 
particles is 62\% at 95\% CL (52\% at 90\% CL).

\begin{figure}[t] 
  \begin{center}
    \includegraphics[width=0.49\textwidth]{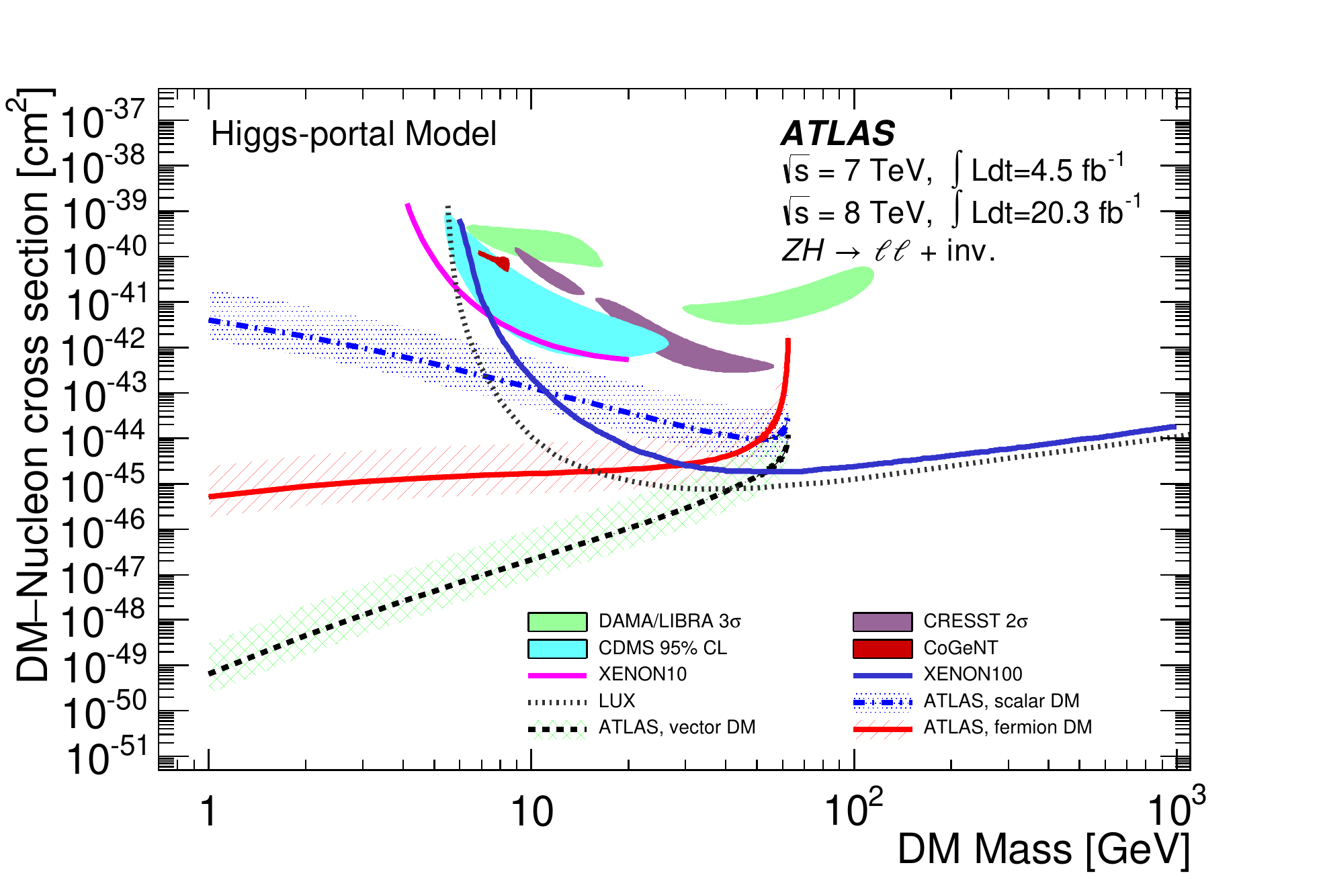}
  \end{center}
\caption{\label{fig:dmlimits} 
Limits on the DM--nucleon scattering cross section at 90\% CL, extracted from the $\text{BR}(H\to \text{inv.})$ limit in a Higgs-portal scenario, compared to results from direct-search experiments~\cite{Bernabei:2008yi,Angle:2011,Aprile:2012,Angloher:2011uu,Aalseth:2011wp,Fox:2011px,Agnese:2013,Akerib:2013}. Cross-section limits and favored regions correspond to a 90\% CL, unless stated otherwise in the legend. Favored regions for DAMA and CoGeNT are based on Ref.~\cite{Fox:2011px}. The results from the direct-search experiments do not depend on the assumptions of the Higgs-portal scenario. 
}
\end{figure}

Within the context of a Higgs-portal DM scenario~\cite{higgs_portal}, in which the Higgs 
boson acts as the mediator particle between DM and SM particles, 
the Higgs boson can decay to a pair of DM particles. 
In this case the limit on  $\text{BR}(H\to \text{inv.})$ for the 125.5 GeV Higgs boson 
can be interpreted in terms of 
an upper limit on the DM--nucleon scattering cross section~\cite{DM_LHC}.
The formalism used to interpret the $\text{BR}(H\to \text{inv.})$ limit in terms of the spin-independent 
DM--nucleon scattering cross sections is described in Refs.~\cite{Kanemura:2010sh,Djouadi:2011aa}.
Figure~\ref{fig:dmlimits} shows 90\% CL upper limits on the DM--nucleon scattering cross section for three model variants 
in which a single DM candidate is considered and is either a scalar, a vector or a Majorana fermion. 
The Higgs--nucleon coupling 
is taken as $0.33^{+ 0.30}_{- 0.07}$~\cite{Djouadi:2011aa}, 
the uncertainty of which is expressed by the bands in the figure. 
Spin-independent results from direct-search experiments are also 
shown~\cite{Bernabei:2008yi,Angle:2011,Aprile:2012,Angloher:2011uu,Aalseth:2011wp,Fox:2011px,Agnese:2013,Akerib:2013}.
These results do not depend on the assumptions of the Higgs-portal scenario. Within the 
constraints of such a scenario however, the results presented in this Letter 
provide the strongest available limits for low-mass DM candidates. 
There is no sensitivity to these models once the mass of the DM candidate exceeds $m_H/2$. 
A search by the ATLAS experiment for DM in more generic models, also using the 
dilepton~$+$~large \met\, final state, is presented in Ref.~\cite{atlas_monoZ}.

We thank CERN for the very successful operation of the LHC, as well as the
support staff from our institutions without whom ATLAS could not be
operated efficiently.

We acknowledge the support of ANPCyT, Argentina; YerPhI, Armenia; ARC,
Australia; BMWF and FWF, Austria; ANAS, Azerbaijan; SSTC, Belarus; CNPq and FAPESP,
Brazil; NSERC, NRC and CFI, Canada; CERN; CONICYT, Chile; CAS, MOST and NSFC,
China; COLCIENCIAS, Colombia; MSMT CR, MPO CR and VSC CR, Czech Republic;
DNRF, DNSRC and Lundbeck Foundation, Denmark; EPLANET, ERC and NSRF, European Union;
IN2P3-CNRS, CEA-DSM/IRFU, France; GNSF, Georgia; BMBF, DFG, HGF, MPG and AvH
Foundation, Germany; GSRT and NSRF, Greece; ISF, MINERVA, GIF, DIP and Benoziyo Center,
Israel; INFN, Italy; MEXT and JSPS, Japan; CNRST, Morocco; FOM and NWO,
Netherlands; BRF and RCN, Norway; MNiSW and NCN, Poland; GRICES and FCT, Portugal; MNE/IFA, Romania; MES of Russia and ROSATOM, Russian Federation; JINR; MSTD,
Serbia; MSSR, Slovakia; ARRS and MIZ\v{S}, Slovenia; DST/NRF, South Africa;
MINECO, Spain; SRC and Wallenberg Foundation, Sweden; SER, SNSF and Cantons of
Bern and Geneva, Switzerland; NSC, Taiwan; TAEK, Turkey; STFC, the Royal
Society and Leverhulme Trust, United Kingdom; DOE and NSF, United States of
America.

The crucial computing support from all WLCG partners is acknowledged
gratefully, in particular from CERN and the ATLAS Tier-1 facilities at
TRIUMF (Canada), NDGF (Denmark, Norway, Sweden), CC-IN2P3 (France),
KIT/GridKA (Germany), INFN-CNAF (Italy), NL-T1 (Netherlands), PIC (Spain),
ASGC (Taiwan), RAL (UK) and BNL (USA) and in the Tier-2 facilities
worldwide.

\bibliography{zhinv}

\begin{thebibliography}{74}%
\makeatletter
\providecommand \@ifxundefined [1]{%
 \@ifx{#1\undefined}
}%
\providecommand \@ifnum [1]{%
 \ifnum #1\expandafter \@firstoftwo
 \else \expandafter \@secondoftwo
 \fi
}%
\providecommand \@ifx [1]{%
 \ifx #1\expandafter \@firstoftwo
 \else \expandafter \@secondoftwo
 \fi
}%
\providecommand \natexlab [1]{#1}%
\providecommand \enquote  [1]{``#1''}%
\providecommand \bibnamefont  [1]{#1}%
\providecommand \bibfnamefont [1]{#1}%
\providecommand \citenamefont [1]{#1}%
\providecommand \href@noop [0]{\@secondoftwo}%
\providecommand \href [0]{\begingroup \@sanitize@url \@href}%
\providecommand \@href[1]{\@@startlink{#1}\@@href}%
\providecommand \@@href[1]{\endgroup#1\@@endlink}%
\providecommand \@sanitize@url [0]{\catcode `\\12\catcode `\$12\catcode
  `\&12\catcode `\#12\catcode `\^12\catcode `\_12\catcode `\%12\relax}%
\providecommand \@@startlink[1]{}%
\providecommand \@@endlink[0]{}%
\providecommand \url  [0]{\begingroup\@sanitize@url \@url }%
\providecommand \@url [1]{\endgroup\@href {#1}{\urlprefix }}%
\providecommand \urlprefix  [0]{URL }%
\providecommand \Eprint [0]{\href }%
\providecommand \doibase [0]{http://dx.doi.org/}%
\providecommand \selectlanguage [0]{\@gobble}%
\providecommand \bibinfo  [0]{\@secondoftwo}%
\providecommand \bibfield  [0]{\@secondoftwo}%
\providecommand \translation [1]{[#1]}%
\providecommand \BibitemOpen [0]{}%
\providecommand \bibitemStop [0]{}%
\providecommand \bibitemNoStop [0]{.\EOS\space}%
\providecommand \EOS [0]{\spacefactor3000\relax}%
\providecommand \BibitemShut  [1]{\csname bibitem#1\endcsname}%
\let\auto@bib@innerbib\@empty
\bibitem [{\citenamefont {Englert}\ and\ \citenamefont
  {Brout}(1964)}]{Englert:1964et}%
  \BibitemOpen
  \bibfield  {author} {\bibinfo {author} {\bibfnamefont {F.}~\bibnamefont
  {Englert}}\ and\ \bibinfo {author} {\bibfnamefont {R.}~\bibnamefont
  {Brout}},\ }\href@noop {} {\bibfield  {journal} {\bibinfo  {journal} {Phys.
  Rev. Lett.}\ }\textbf {\bibinfo {volume} {13}},\ \bibinfo {pages} {321}
  (\bibinfo {year} {1964})}\BibitemShut {NoStop}%
\bibitem [{\citenamefont {Higgs}(1964)}]{Higgs:1964pj}%
  \BibitemOpen
  \bibfield  {author} {\bibinfo {author} {\bibfnamefont {P.~W.}\ \bibnamefont
  {Higgs}},\ }\href@noop {} {\bibfield  {journal} {\bibinfo  {journal} {Phys.
  Rev. Lett.}\ }\textbf {\bibinfo {volume} {13}},\ \bibinfo {pages} {508}
  (\bibinfo {year} {1964})}\BibitemShut {NoStop}%
\bibitem [{\citenamefont {Guralnik}\ \emph {et~al.}(1964)\citenamefont
  {Guralnik}, \citenamefont {Hagen},\ and\ \citenamefont
  {Kibble}}]{Guralnik:1964eu}%
  \BibitemOpen
  \bibfield  {author} {\bibinfo {author} {\bibfnamefont {G.}~\bibnamefont
  {Guralnik}}, \bibinfo {author} {\bibfnamefont {C.}~\bibnamefont {Hagen}}, \
  and\ \bibinfo {author} {\bibfnamefont {T.}~\bibnamefont {Kibble}},\
  }\href@noop {} {\bibfield  {journal} {\bibinfo  {journal} {Phys. Rev. Lett.}\
  }\textbf {\bibinfo {volume} {13}},\ \bibinfo {pages} {585} (\bibinfo {year}
  {1964})}\BibitemShut {NoStop}%
\bibitem [{\citenamefont {{K. Griest and H. E.
  Haber}}(1988)}]{Hinv:Griest1988}%
  \BibitemOpen
  \bibfield  {author} {\bibinfo {author} {\bibnamefont {{K. Griest and H. E.
  Haber}}},\ }\href@noop {} {\bibfield  {journal} {\bibinfo  {journal} {Phys.
  Rev. D}\ }\textbf {\bibinfo {volume} {37}},\ \bibinfo {pages} {719} (\bibinfo
  {year} {1988})}\BibitemShut {NoStop}%
\bibitem [{\citenamefont {{A. Djouadi, P. Janot, J. Kalinowski, and P. M.
  Zerwas}}(1996)}]{Hinv:Djouadi1996}%
  \BibitemOpen
  \bibfield  {author} {\bibinfo {author} {\bibnamefont {{A. Djouadi, P. Janot,
  J. Kalinowski, and P. M. Zerwas}}},\ }\href@noop {} {\bibfield  {journal}
  {\bibinfo  {journal} {Phys. Lett. B}\ }\textbf {\bibinfo {volume} {376}},\
  \bibinfo {pages} {220} (\bibinfo {year} {1996})},\ \Eprint
  {http://arxiv.org/abs/hep-ph/9603368} {arXiv:hep-ph/9603368} \BibitemShut
  {NoStop}%
\bibitem [{\citenamefont {{I. Antoniadis, M. Tuckmantel, and F.
  Zwirner}}(2005)}]{Hinv:Antoniadis2005}%
  \BibitemOpen
  \bibfield  {author} {\bibinfo {author} {\bibnamefont {{I. Antoniadis, M.
  Tuckmantel, and F. Zwirner}}},\ }\href@noop {} {\bibfield  {journal}
  {\bibinfo  {journal} {Nucl. Phys. B}\ }\textbf {\bibinfo {volume} {707}},\
  \bibinfo {pages} {215} (\bibinfo {year} {2005})},\ \Eprint
  {http://arxiv.org/abs/hep-ph/0410165} {arXiv:hep-ph/0410165} \BibitemShut
  {NoStop}%
\bibitem [{\citenamefont {{T. Binoth and J. J. van der
  Bij}}(1997)}]{Hinv:Binoth1997}%
  \BibitemOpen
  \bibfield  {author} {\bibinfo {author} {\bibnamefont {{T. Binoth and J. J.
  van der Bij}}},\ }\href@noop {} {\bibfield  {journal} {\bibinfo  {journal}
  {Z. Phys. C}\ }\textbf {\bibinfo {volume} {75}},\ \bibinfo {pages} {17}
  (\bibinfo {year} {1997})},\ \Eprint {http://arxiv.org/abs/hep-ph/9608245}
  {arXiv:hep-ph/9608245} \BibitemShut {NoStop}%
\bibitem [{\citenamefont {{Y. Chikashige, R. N. Mohapatra, and R. D.
  Peccei}}(1981)}]{Hinv:Chikashige1981}%
  \BibitemOpen
  \bibfield  {author} {\bibinfo {author} {\bibnamefont {{Y. Chikashige, R. N.
  Mohapatra, and R. D. Peccei}}},\ }\href@noop {} {\bibfield  {journal}
  {\bibinfo  {journal} {Phys. Lett. B}\ }\textbf {\bibinfo {volume} {98}},\
  \bibinfo {pages} {265} (\bibinfo {year} {1981})}\BibitemShut {NoStop}%
\bibitem [{\citenamefont {{A. S. Joshipura and S. D.
  Rindani}}(1992)}]{Hinv:Joshipura1992}%
  \BibitemOpen
  \bibfield  {author} {\bibinfo {author} {\bibnamefont {{A. S. Joshipura and S.
  D. Rindani}}},\ }\href@noop {} {\bibfield  {journal} {\bibinfo  {journal}
  {Phys. Rev. Lett.}\ }\textbf {\bibinfo {volume} {69}},\ \bibinfo {pages}
  {3269} (\bibinfo {year} {1992})}\BibitemShut {NoStop}%
\bibitem [{\citenamefont {{R. E. Shrock and M.
  Suzuki}}(1982)}]{Hinv:Shrock1982}%
  \BibitemOpen
  \bibfield  {author} {\bibinfo {author} {\bibnamefont {{R. E. Shrock and M.
  Suzuki}}},\ }\href@noop {} {\bibfield  {journal} {\bibinfo  {journal} {Phys.
  Lett. B}\ }\textbf {\bibinfo {volume} {110}},\ \bibinfo {pages} {250}
  (\bibinfo {year} {1982})}\BibitemShut {NoStop}%
\bibitem [{\citenamefont {{R. N. Mohapatra and J. W. F.
  Valle}}(1986)}]{Hinv:Mohapatra1986}%
  \BibitemOpen
  \bibfield  {author} {\bibinfo {author} {\bibnamefont {{R. N. Mohapatra and J.
  W. F. Valle}}},\ }\href@noop {} {\bibfield  {journal} {\bibinfo  {journal}
  {Phys. Rev. D}\ }\textbf {\bibinfo {volume} {34}},\ \bibinfo {pages} {1642}
  (\bibinfo {year} {1986})}\BibitemShut {NoStop}%
\bibitem [{\citenamefont {{M. C. Gonzalez-Garcia and J. W. F.
  Valle}}(1989)}]{Hinv:Gonzalez1989}%
  \BibitemOpen
  \bibfield  {author} {\bibinfo {author} {\bibnamefont {{M. C. Gonzalez-Garcia
  and J. W. F. Valle}}},\ }\href@noop {} {\bibfield  {journal} {\bibinfo
  {journal} {Phys. Lett. B}\ }\textbf {\bibinfo {volume} {216}},\ \bibinfo
  {pages} {360} (\bibinfo {year} {1989})}\BibitemShut {NoStop}%
\bibitem [{\citenamefont {{E. D. Carlson and L. J.
  Hall}}(1989)}]{Hinv:Carlson1989}%
  \BibitemOpen
  \bibfield  {author} {\bibinfo {author} {\bibnamefont {{E. D. Carlson and L.
  J. Hall}}},\ }\href@noop {} {\bibfield  {journal} {\bibinfo  {journal} {Phys.
  Rev. D}\ }\textbf {\bibinfo {volume} {40}},\ \bibinfo {pages} {3187}
  (\bibinfo {year} {1989})}\BibitemShut {NoStop}%
\bibitem [{\citenamefont {{L. F. Li, Y. Liu, and L.
  Wolfenstein}}(1985)}]{Hinv:Li1985}%
  \BibitemOpen
  \bibfield  {author} {\bibinfo {author} {\bibnamefont {{L. F. Li, Y. Liu, and
  L. Wolfenstein}}},\ }\href@noop {} {\bibfield  {journal} {\bibinfo  {journal}
  {Phys. Lett. B}\ }\textbf {\bibinfo {volume} {159}},\ \bibinfo {pages} {45}
  (\bibinfo {year} {1985})}\BibitemShut {NoStop}%
\bibitem [{\citenamefont {{A. Zee}}(1980)}]{Hinv:Zee1980}%
  \BibitemOpen
  \bibfield  {author} {\bibinfo {author} {\bibnamefont {{A. Zee}}},\
  }\href@noop {} {\bibfield  {journal} {\bibinfo  {journal} {Phys. Lett. B}\
  }\textbf {\bibinfo {volume} {93}},\ \bibinfo {pages} {389} (\bibinfo {year}
  {1980})}\BibitemShut {NoStop}%
\bibitem [{\citenamefont {{F. de Campos, O. J. P. \'Eboli, J. Rosiek, and J. W.
  F. Valle}}(1997)}]{Hinv:Campos1997}%
  \BibitemOpen
  \bibfield  {author} {\bibinfo {author} {\bibnamefont {{F. de Campos, O. J. P.
  \'Eboli, J. Rosiek, and J. W. F. Valle}}},\ }\href@noop {} {\bibfield
  {journal} {\bibinfo  {journal} {Phys. Rev. D}\ }\textbf {\bibinfo {volume}
  {55}},\ \bibinfo {pages} {1316} (\bibinfo {year} {1997})},\ \Eprint
  {http://arxiv.org/abs/hep-ph/9601269} {arXiv:hep-ph/9601269} \BibitemShut
  {NoStop}%
\bibitem [{\citenamefont {{S. P. Martin and J. D.
  Wells}}(1999)}]{Hinv:Martin1999}%
  \BibitemOpen
  \bibfield  {author} {\bibinfo {author} {\bibnamefont {{S. P. Martin and J. D.
  Wells}}},\ }\href@noop {} {\bibfield  {journal} {\bibinfo  {journal} {Phys.
  Rev. D}\ }\textbf {\bibinfo {volume} {60}},\ \bibinfo {pages} {035006}
  (\bibinfo {year} {1999})},\ \Eprint {http://arxiv.org/abs/hep-ph/9903259}
  {arXiv:hep-ph/9903259} \BibitemShut {NoStop}%
\bibitem [{\citenamefont {{G. F. Giudice, R. Rattazzi, and J. D.
  Wells}}(2001)}]{Hinv:Giudice2001}%
  \BibitemOpen
  \bibfield  {author} {\bibinfo {author} {\bibnamefont {{G. F. Giudice, R.
  Rattazzi, and J. D. Wells}}},\ }\href@noop {} {\bibfield  {journal} {\bibinfo
   {journal} {Nucl. Phys. B}\ }\textbf {\bibinfo {volume} {595}},\ \bibinfo
  {pages} {250} (\bibinfo {year} {2001})},\ \Eprint
  {http://arxiv.org/abs/hep-ph/0002178} {arXiv:hep-ph/0002178} \BibitemShut
  {NoStop}%
\bibitem [{\citenamefont {{LEP Higgs Working Group}}(2001)}]{LEP}%
  \BibitemOpen
  \bibfield  {author} {\bibinfo {author} {\bibnamefont {{LEP Higgs Working
  Group}}},\ }\href@noop {} {\bibfield  {journal} {\bibinfo  {journal} {LHWG
  Note 2001-06}\ } (\bibinfo {year} {2001})},\ \Eprint
  {http://arxiv.org/abs/arXiv:hep-ex/0107032} {arXiv:hep-ex/0107032}
  \BibitemShut {NoStop}%
\bibitem [{\citenamefont {{D. Choudhury, and D.
  Roy}}(1994)}]{Hinv:Choudhury1994}%
  \BibitemOpen
  \bibfield  {author} {\bibinfo {author} {\bibnamefont {{D. Choudhury, and D.
  Roy}}},\ }\href@noop {} {\bibfield  {journal} {\bibinfo  {journal} {Phys.
  Lett. B}\ }\textbf {\bibinfo {volume} {322}},\ \bibinfo {pages} {368}
  (\bibinfo {year} {1994})},\ \Eprint {http://arxiv.org/abs/hep-ph/9312347}
  {arXiv:hep-ph/9312347} \BibitemShut {NoStop}%
\bibitem [{\citenamefont {{R. M. Godbole, M. Guchait, K. Mazumdar, S. Moretti,
  and D.P. Roy}}(2003)}]{Hinv:Godbole2003}%
  \BibitemOpen
  \bibfield  {author} {\bibinfo {author} {\bibnamefont {{R. M. Godbole, M.
  Guchait, K. Mazumdar, S. Moretti, and D.P. Roy}}},\ }\href@noop {} {\bibfield
   {journal} {\bibinfo  {journal} {Phys. Lett. B}\ }\textbf {\bibinfo {volume}
  {571}},\ \bibinfo {pages} {184} (\bibinfo {year} {2003})},\ \Eprint
  {http://arxiv.org/abs/hep-ph/0304137} {arXiv:hep-ph/0304137} \BibitemShut
  {NoStop}%
\bibitem [{\citenamefont {{H. Davoudiasl, T. Han, H. E.
  Logan}}(2005)}]{Hinv:Davoudiasl2005}%
  \BibitemOpen
  \bibfield  {author} {\bibinfo {author} {\bibnamefont {{H. Davoudiasl, T. Han,
  H. E. Logan}}},\ }\href@noop {} {\bibfield  {journal} {\bibinfo  {journal}
  {Phys. Rev. D}\ }\textbf {\bibinfo {volume} {71}},\ \bibinfo {pages} {115007}
  (\bibinfo {year} {2005})},\ \Eprint {http://arxiv.org/abs/hep-ph/0412269}
  {arXiv:hep-ph/0412269} \BibitemShut {NoStop}%
\bibitem [{\citenamefont {{ATLAS
  Collaboration}}(2013{\natexlab{a}})}]{Aad:2013wqa}%
  \BibitemOpen
  \bibfield  {author} {\bibinfo {author} {\bibnamefont {{ATLAS
  Collaboration}}},\ }\href@noop {} {\bibfield  {journal} {\bibinfo  {journal}
  {Phys. Lett. B}\ }\textbf {\bibinfo {volume} {726}},\ \bibinfo {pages} {88}
  (\bibinfo {year} {2013}{\natexlab{a}})},\ \Eprint
  {http://arxiv.org/abs/1307.1427} {arXiv:1307.1427} \BibitemShut {NoStop}%
\bibitem [{\citenamefont {{O. Brein, A. Djouadi and R.
  Harlander}}(2004)}]{Brein:2003wg}%
  \BibitemOpen
  \bibfield  {author} {\bibinfo {author} {\bibnamefont {{O. Brein, A. Djouadi
  and R. Harlander}}},\ }\href@noop {} {\bibfield  {journal} {\bibinfo
  {journal} {Phys. Lett. B}\ }\textbf {\bibinfo {volume} {579}},\ \bibinfo
  {pages} {149} (\bibinfo {year} {2004})},\ \Eprint
  {http://arxiv.org/abs/hep-ph/0307206} {arXiv:hep-ph/0307206} \BibitemShut
  {NoStop}%
\bibitem [{\citenamefont {{M. L. Ciccolini, S. Dittmaier and M.
  Kr\"amer}}(2003)}]{Ciccolini:2003jy}%
  \BibitemOpen
  \bibfield  {author} {\bibinfo {author} {\bibnamefont {{M. L. Ciccolini, S.
  Dittmaier and M. Kr\"amer}}},\ }\href@noop {} {\bibfield  {journal} {\bibinfo
   {journal} {Phys. Rev. D}\ }\textbf {\bibinfo {volume} {68}},\ \bibinfo
  {pages} {073003} (\bibinfo {year} {2003})},\ \Eprint
  {http://arxiv.org/abs/hep-ph/0306234} {arXiv:hep-ph/0306234} \BibitemShut
  {NoStop}%
\bibitem [{\citenamefont {{A. Denner, S. Dittmaier, S. Kallweit and A.
  M\"uck}}(2012)}]{Denner:2011id}%
  \BibitemOpen
  \bibfield  {author} {\bibinfo {author} {\bibnamefont {{A. Denner, S.
  Dittmaier, S. Kallweit and A. M\"uck}}},\ }\href@noop {} {\bibfield
  {journal} {\bibinfo  {journal} {J. High Energy Phys.}\ }\textbf {\bibinfo
  {volume} {{\rm 03}}},\ \bibinfo {pages} {075} (\bibinfo {year} {2012})},\
  \Eprint {http://arxiv.org/abs/1112.5142} {arXiv:1112.5142} \BibitemShut
  {NoStop}%
\bibitem [{\citenamefont {{LHC Higgs Cross Section Working Group}}\ \emph
  {et~al.}(2013)\citenamefont {{LHC Higgs Cross Section Working Group}},
  \citenamefont {Dittmaier}, \citenamefont {Mariotti}, \citenamefont
  {Passarino},\ and\ \citenamefont {Tanaka~(Eds.)}}]{LHCHiggsYR3}%
  \BibitemOpen
  \bibfield  {author} {\bibinfo {author} {\bibnamefont {{LHC Higgs Cross
  Section Working Group}}}, \bibinfo {author} {\bibfnamefont {S.}~\bibnamefont
  {Dittmaier}}, \bibinfo {author} {\bibfnamefont {C.}~\bibnamefont {Mariotti}},
  \bibinfo {author} {\bibfnamefont {G.}~\bibnamefont {Passarino}}, \ and\
  \bibinfo {author} {\bibfnamefont {R.}~\bibnamefont {Tanaka~(Eds.)}},\
  }\href@noop {} {\bibfield  {journal} {\bibinfo  {journal} {CERN-2013-004}\ }
  (\bibinfo {year} {2013})},\ \Eprint {http://arxiv.org/abs/1307.1347}
  {arXiv:1307.1347} \BibitemShut {NoStop}%
\bibitem [{\citenamefont {{ATLAS Collaboration}}(2008)}]{Aad:2008zzm}%
  \BibitemOpen
  \bibfield  {author} {\bibinfo {author} {\bibnamefont {{ATLAS
  Collaboration}}},\ }\href@noop {} {\bibfield  {journal} {\bibinfo  {journal}
  {JINST}\ }\textbf {\bibinfo {volume} {3}},\ \bibinfo {pages} {S08003}
  (\bibinfo {year} {2008})}\BibitemShut {NoStop}%
\bibitem [{\citenamefont {{S. Agostinelli {\it et al.} (GEANT4
  Collaboration)}}(2003)}]{geant4}%
  \BibitemOpen
  \bibfield  {author} {\bibinfo {author} {\bibnamefont {{S. Agostinelli {\it et
  al.} (GEANT4 Collaboration)}}},\ }\href@noop {} {\bibfield  {journal}
  {\bibinfo  {journal} {Nucl. Instrum. Methods Phys. Res., Sect. A}\ }\textbf
  {\bibinfo {volume} {506}},\ \bibinfo {pages} {250} (\bibinfo {year}
  {2003})}\BibitemShut {NoStop}%
\bibitem [{\citenamefont {{ATLAS
  Collaboration}}(2010{\natexlab{a}})}]{simuAtlas}%
  \BibitemOpen
  \bibfield  {author} {\bibinfo {author} {\bibnamefont {{ATLAS
  Collaboration}}},\ }\href@noop {} {\bibfield  {journal} {\bibinfo  {journal}
  {Eur.~Phys.~J.~C}\ }\textbf {\bibinfo {volume} {70}},\ \bibinfo {pages} {823}
  (\bibinfo {year} {2010}{\natexlab{a}})},\ \Eprint
  {http://arxiv.org/abs/1005.4568} {arXiv:1005.4568} \BibitemShut {NoStop}%
\bibitem [{\citenamefont {{M. B\"ahr {\it et al.}}}(2008)}]{herwigpp}%
  \BibitemOpen
  \bibfield  {author} {\bibinfo {author} {\bibnamefont {{M. B\"ahr {\it et
  al.}}}},\ }\href@noop {} {\bibfield  {journal} {\bibinfo  {journal} {Eur.
  Phys. J. C}\ }\textbf {\bibinfo {volume} {58}},\ \bibinfo {pages} {639}
  (\bibinfo {year} {2008})},\ \Eprint {http://arxiv.org/abs/0803.0883}
  {arXiv:0803.0883} \BibitemShut {NoStop}%
\bibitem [{\citenamefont {{P. Nason}}(2004)}]{powheg1}%
  \BibitemOpen
  \bibfield  {author} {\bibinfo {author} {\bibnamefont {{P. Nason}}},\
  }\href@noop {} {\bibfield  {journal} {\bibinfo  {journal} {J. High Energy
  Phys.}\ }\textbf {\bibinfo {volume} {{\rm 11}}},\ \bibinfo {pages} {040}
  (\bibinfo {year} {2004})},\ \Eprint
  {http://arxiv.org/abs/arXiv:hep-ph/0409146} {arXiv:hep-ph/0409146}
  \BibitemShut {NoStop}%
\bibitem [{\citenamefont {{S. Frixione, P. Nason and C.
  Oleari}}(2007)}]{powheg2}%
  \BibitemOpen
  \bibfield  {author} {\bibinfo {author} {\bibnamefont {{S. Frixione, P. Nason
  and C. Oleari}}},\ }\href@noop {} {\bibfield  {journal} {\bibinfo  {journal}
  {J. High Energy Phys.}\ }\textbf {\bibinfo {volume} {{\rm 11}}},\ \bibinfo
  {pages} {070} (\bibinfo {year} {2007})},\ \Eprint
  {http://arxiv.org/abs/0709.2092} {arXiv:0709.2092} \BibitemShut {NoStop}%
\bibitem [{\citenamefont {{S. Alioli, P. Nason, C. Oleari and E.
  Re}}(2010)}]{powheg3}%
  \BibitemOpen
  \bibfield  {author} {\bibinfo {author} {\bibnamefont {{S. Alioli, P. Nason,
  C. Oleari and E. Re}}},\ }\href@noop {} {\bibfield  {journal} {\bibinfo
  {journal} {J. High Energy Phys.}\ }\textbf {\bibinfo {volume} {{\rm 06}}},\
  \bibinfo {pages} {043} (\bibinfo {year} {2010})},\ \Eprint
  {http://arxiv.org/abs/1002.2581} {arXiv:1002.2581} \BibitemShut {NoStop}%
\bibitem [{\citenamefont {{T. Sj\"ostrand, S. Mrenna and P.
  Skands}}(2008)}]{Sjostrand:2007gs}%
  \BibitemOpen
  \bibfield  {author} {\bibinfo {author} {\bibnamefont {{T. Sj\"ostrand, S.
  Mrenna and P. Skands}}},\ }\href@noop {} {\bibfield  {journal} {\bibinfo
  {journal} {Comput. Phys. Commun.}\ }\textbf {\bibinfo {volume} {178}},\
  \bibinfo {pages} {852} (\bibinfo {year} {2008})},\ \Eprint
  {http://arxiv.org/abs/0710.3820} {arXiv:0710.3820} \BibitemShut {NoStop}%
\bibitem [{\citenamefont {{T. Melia, P. Nason, R. R\"ontsch, and G.
  Zanderighi}}(2011)}]{powheg_ew}%
  \BibitemOpen
  \bibfield  {author} {\bibinfo {author} {\bibnamefont {{T. Melia, P. Nason, R.
  R\"ontsch, and G. Zanderighi}}},\ }\href@noop {} {\bibfield  {journal}
  {\bibinfo  {journal} {J. High Energy Phys.}\ }\textbf {\bibinfo {volume}
  {{\rm 11}}},\ \bibinfo {pages} {078} (\bibinfo {year} {2011})},\ \Eprint
  {http://arxiv.org/abs/1107.5051} {arXiv:1107.5051} \BibitemShut {NoStop}%
\bibitem [{\citenamefont {{G. Corcella, I. G. Knowles, G. Marchesini, S.
  Moretti, K. Odagiri, P. Richardson, M. H. Seymour, B. R.
  Webber}}(2001)}]{herwig}%
  \BibitemOpen
  \bibfield  {author} {\bibinfo {author} {\bibnamefont {{G. Corcella, I. G.
  Knowles, G. Marchesini, S. Moretti, K. Odagiri, P. Richardson, M. H. Seymour,
  B. R. Webber}}},\ }\href@noop {} {\bibfield  {journal} {\bibinfo  {journal}
  {J. High Energy Phys.}\ }\textbf {\bibinfo {volume} {{\rm 01}}},\ \bibinfo
  {pages} {010} (\bibinfo {year} {2001})},\ \Eprint
  {http://arxiv.org/abs/arXiv:hep-ph/0011363} {arXiv:hep-ph/0011363}
  \BibitemShut {NoStop}%
\bibitem [{\citenamefont {{T. Gleisberg, S. H\"oche, F. Krauss, M. Sch\"onherr,
  S. Schumann, F. Siegert, J. Winter}}(2009)}]{sherpa}%
  \BibitemOpen
  \bibfield  {author} {\bibinfo {author} {\bibnamefont {{T. Gleisberg, S.
  H\"oche, F. Krauss, M. Sch\"onherr, S. Schumann, F. Siegert, J. Winter}}},\
  }\href@noop {} {\bibfield  {journal} {\bibinfo  {journal} {J. High Energy
  Phys.}\ }\textbf {\bibinfo {volume} {{\rm 02}}},\ \bibinfo {pages} {007}
  (\bibinfo {year} {2009})},\ \Eprint {http://arxiv.org/abs/arXiv:0811.4622}
  {arXiv:0811.4622} \BibitemShut {NoStop}%
\bibitem [{\citenamefont {{T. Binoth, M. Ciccolini, N. Kauer, and M.
  Kr\"amer}}(2006)}]{gg2ww}%
  \BibitemOpen
  \bibfield  {author} {\bibinfo {author} {\bibnamefont {{T. Binoth, M.
  Ciccolini, N. Kauer, and M. Kr\"amer}}},\ }\href@noop {} {\bibfield
  {journal} {\bibinfo  {journal} {J. High Energy Phys.}\ }\textbf {\bibinfo
  {volume} {{\rm 12}}},\ \bibinfo {pages} {046} (\bibinfo {year} {2006})},\
  \Eprint {http://arxiv.org/abs/hep-ph/0611170} {arXiv:hep-ph/0611170}
  \BibitemShut {NoStop}%
\bibitem [{\citenamefont {{J.~M.~Butterworth, J.~R.~Forshaw and
  M.~H.~Seymour}}(1996)}]{jimmy}%
  \BibitemOpen
  \bibfield  {author} {\bibinfo {author} {\bibnamefont {{J.~M.~Butterworth,
  J.~R.~Forshaw and M.~H.~Seymour}}},\ }\href@noop {} {\bibfield  {journal}
  {\bibinfo  {journal} {Z. Phys. C}\ }\textbf {\bibinfo {volume} {72}},\
  \bibinfo {pages} {637} (\bibinfo {year} {1996})},\ \Eprint
  {http://arxiv.org/abs/hep-ph/9601371} {arXiv:hep-ph/9601371} \BibitemShut
  {NoStop}%
\bibitem [{\citenamefont {{S.~Frixione and B.~R.~Webber}}(2002)}]{mcatnlo}%
  \BibitemOpen
  \bibfield  {author} {\bibinfo {author} {\bibnamefont {{S.~Frixione and
  B.~R.~Webber}}},\ }\href@noop {} {\bibfield  {journal} {\bibinfo  {journal}
  {J. High Energy Phys.}\ }\textbf {\bibinfo {volume} {{\rm 06}}},\ \bibinfo
  {pages} {029} (\bibinfo {year} {2002})},\ \Eprint
  {http://arxiv.org/abs/hep-ph/0204244} {arXiv:hep-ph/0204244} \BibitemShut
  {NoStop}%
\bibitem [{\citenamefont {Kersevan}\ and\ \citenamefont
  {Richter-Was}(2013)}]{acermc}%
  \BibitemOpen
  \bibfield  {author} {\bibinfo {author} {\bibfnamefont {B.~P.}\ \bibnamefont
  {Kersevan}}\ and\ \bibinfo {author} {\bibfnamefont {E.}~\bibnamefont
  {Richter-Was}},\ }\href@noop {} {\bibfield  {journal} {\bibinfo  {journal}
  {Comput. Phys. Commun.}\ }\textbf {\bibinfo {volume} {184}},\ \bibinfo
  {pages} {919 } (\bibinfo {year} {2013})},\ \Eprint
  {http://arxiv.org/abs/hep-ph/0405247} {arXiv:hep-ph/0405247} \BibitemShut
  {NoStop}%
\bibitem [{\citenamefont {{T.~Sj\"ostrand, S.~Mrenna and
  P.~Skands}}(2006)}]{pythia}%
  \BibitemOpen
  \bibfield  {author} {\bibinfo {author} {\bibnamefont {{T.~Sj\"ostrand,
  S.~Mrenna and P.~Skands}}},\ }\href@noop {} {\bibfield  {journal} {\bibinfo
  {journal} {J. High Energy Phys.}\ }\textbf {\bibinfo {volume} {{\rm 05}}},\
  \bibinfo {pages} {026} (\bibinfo {year} {2006})},\ \Eprint
  {http://arxiv.org/abs/arXiv:hep-ph/0603175} {arXiv:hep-ph/0603175}
  \BibitemShut {NoStop}%
\bibitem [{\citenamefont {{M.~L.~Mangano, F. Piccinini, A. D. Polosa, M.
  Moretti and R. Pittau}}(2003)}]{alpgen}%
  \BibitemOpen
  \bibfield  {author} {\bibinfo {author} {\bibnamefont {{M.~L.~Mangano, F.
  Piccinini, A. D. Polosa, M. Moretti and R. Pittau}}},\ }\href@noop {}
  {\bibfield  {journal} {\bibinfo  {journal} {J. High Energy Phys.}\ }\textbf
  {\bibinfo {volume} {{\rm 07}}},\ \bibinfo {pages} {001} (\bibinfo {year}
  {2003})},\ \Eprint {http://arxiv.org/abs/arXiv:hep-ph/0206293}
  {arXiv:hep-ph/0206293} \BibitemShut {NoStop}%
\bibitem [{\citenamefont {Alioli}\ \emph {et~al.}(2009)\citenamefont {Alioli},
  \citenamefont {Nason}, \citenamefont {Oleari},\ and\ \citenamefont
  {Re}}]{powheg4}%
  \BibitemOpen
  \bibfield  {author} {\bibinfo {author} {\bibfnamefont {S.}~\bibnamefont
  {Alioli}}, \bibinfo {author} {\bibfnamefont {P.}~\bibnamefont {Nason}},
  \bibinfo {author} {\bibfnamefont {C.}~\bibnamefont {Oleari}}, \ and\ \bibinfo
  {author} {\bibfnamefont {E.}~\bibnamefont {Re}},\ }\href@noop {} {\bibfield
  {journal} {\bibinfo  {journal} {J. High Energy Phys.}\ }\textbf {\bibinfo
  {volume} {{\rm 04}}},\ \bibinfo {pages} {002} (\bibinfo {year} {2009})},\
  \Eprint {http://arxiv.org/abs/0812.0578} {arXiv:0812.0578} \BibitemShut
  {NoStop}%
\bibitem [{\citenamefont {Nason}\ and\ \citenamefont {Oleari}(2010)}]{powheg5}%
  \BibitemOpen
  \bibfield  {author} {\bibinfo {author} {\bibfnamefont {P.}~\bibnamefont
  {Nason}}\ and\ \bibinfo {author} {\bibfnamefont {C.}~\bibnamefont {Oleari}},\
  }\href@noop {} {\bibfield  {journal} {\bibinfo  {journal} {J. High Energy
  Phys.}\ }\textbf {\bibinfo {volume} {{\rm 02}}},\ \bibinfo {pages} {037}
  (\bibinfo {year} {2010})},\ \Eprint {http://arxiv.org/abs/0911.5299}
  {arXiv:0911.5299} \BibitemShut {NoStop}%
\bibitem [{\citenamefont {{ATLAS
  Collaboration}}(2012{\natexlab{a}})}]{Aad:2011mk}%
  \BibitemOpen
  \bibfield  {author} {\bibinfo {author} {\bibnamefont {{ATLAS
  Collaboration}}},\ }\href@noop {} {\bibfield  {journal} {\bibinfo  {journal}
  {Eur. Phys. J. C}\ }\textbf {\bibinfo {volume} {72}},\ \bibinfo {pages}
  {1909} (\bibinfo {year} {2012}{\natexlab{a}})},\ \Eprint
  {http://arxiv.org/abs/1110.3174} {arXiv:1110.3174} \BibitemShut {NoStop}%
\bibitem [{Atl()}]{AtlasCoordinateSystem}%
  \BibitemOpen
  \bibinfo {note} {ATLAS uses a right-handed coordinate system with its origin
  at the nominal interaction point in the center of the detector and the
  $z$-axis along the beam pipe. Polar coordinates ($r$, $\phi$) are used in the
  transverse plane, $\phi$ being the azimuthal angle around the beam pipe. The
  pseudorapidity $\eta$ is defined in terms of the polar angle $\theta$ as
  $\eta = - \ln\, \tan(\theta/2)$, and $\Delta R =
  \sqrt{(\Delta\eta)^2+(\Delta\phi)^2}$.}\BibitemShut {Stop}%
\bibitem [{\citenamefont {{ATLAS
  Collaboration}}(2010{\natexlab{b}})}]{Aad:2010yt}%
  \BibitemOpen
  \bibfield  {author} {\bibinfo {author} {\bibnamefont {{ATLAS
  Collaboration}}},\ }\href@noop {} {\bibfield  {journal} {\bibinfo  {journal}
  {J. High Energy Phys.}\ }\textbf {\bibinfo {volume} {{\rm 12}}},\ \bibinfo
  {pages} {060} (\bibinfo {year} {2010}{\natexlab{b}})},\ \Eprint
  {http://arxiv.org/abs/1010.2130} {arXiv:1010.2130} \BibitemShut {NoStop}%
\bibitem [{\citenamefont {{M.~Cacciari, G.~P.~Salam and
  G.~Soyez}}(2008)}]{antikt}%
  \BibitemOpen
  \bibfield  {author} {\bibinfo {author} {\bibnamefont {{M.~Cacciari,
  G.~P.~Salam and G.~Soyez}}},\ }\href@noop {} {\bibfield  {journal} {\bibinfo
  {journal} {J. High Energy Phys.}\ }\textbf {\bibinfo {volume} {{\rm 04}}},\
  \bibinfo {pages} {063} (\bibinfo {year} {2008})},\ \Eprint
  {http://arxiv.org/abs/0802.1189} {arXiv:0802.1189} \BibitemShut {NoStop}%
\bibitem [{\citenamefont {{ATLAS
  Collaboration}}(2012{\natexlab{b}})}]{metcalc}%
  \BibitemOpen
  \bibfield  {author} {\bibinfo {author} {\bibnamefont {{ATLAS
  Collaboration}}},\ }\href@noop {} {\bibfield  {journal} {\bibinfo  {journal}
  {Eur. Phys. J. C}\ }\textbf {\bibinfo {volume} {72}},\ \bibinfo {pages}
  {1844} (\bibinfo {year} {2012}{\natexlab{b}})},\ \Eprint
  {http://arxiv.org/abs/1108.5602} {arXiv:1108.5602} \BibitemShut {NoStop}%
\bibitem [{\citenamefont {{ATLAS
  Collaboration}}(2010{\natexlab{c}})}]{trackmetCONF}%
  \BibitemOpen
  \bibfield  {author} {\bibinfo {author} {\bibnamefont {{ATLAS
  Collaboration}}},\ }\href {https://cds.cern.ch/record/1277652} {\bibfield
  {journal} {\bibinfo  {journal} {ATLAS-CONF-2010-020}\ } (\bibinfo {year}
  {2010}{\natexlab{c}})}\BibitemShut {NoStop}%
\bibitem [{\citenamefont {{ATLAS
  Collaboration}}(2013{\natexlab{b}})}]{Aad:2013zz}%
  \BibitemOpen
  \bibfield  {author} {\bibinfo {author} {\bibnamefont {{ATLAS
  Collaboration}}},\ }\href@noop {} {\bibfield  {journal} {\bibinfo  {journal}
  {J. High Energy Phys.}\ }\textbf {\bibinfo {volume} {{\rm 03}}},\ \bibinfo
  {pages} {128} (\bibinfo {year} {2013}{\natexlab{b}})},\ \Eprint
  {http://arxiv.org/abs/1211.6096} {arXiv:1211.6096} \BibitemShut {NoStop}%
\bibitem [{\citenamefont {{ATLAS
  Collaboration}}(2011{\natexlab{a}})}]{Aad:2010ey}%
  \BibitemOpen
  \bibfield  {author} {\bibinfo {author} {\bibnamefont {{ATLAS
  Collaboration}}},\ }\href@noop {} {\bibfield  {journal} {\bibinfo  {journal}
  {Eur. Phys. J. C}\ }\textbf {\bibinfo {volume} {71}},\ \bibinfo {pages}
  {1577} (\bibinfo {year} {2011}{\natexlab{a}})},\ \Eprint
  {http://arxiv.org/abs/1012.1792} {arXiv:1012.1792} \BibitemShut {NoStop}%
\bibitem [{\citenamefont {{ATLAS
  Collaboration}}(2011{\natexlab{b}})}]{Aad:2011kk}%
  \BibitemOpen
  \bibfield  {author} {\bibinfo {author} {\bibnamefont {{ATLAS
  Collaboration}}},\ }\href@noop {} {\bibfield  {journal} {\bibinfo  {journal}
  {Phys. Rev. Lett.}\ }\textbf {\bibinfo {volume} {107}},\ \bibinfo {pages}
  {041802} (\bibinfo {year} {2011}{\natexlab{b}})},\ \Eprint
  {http://arxiv.org/abs/1104.5225} {arXiv:1104.5225} \BibitemShut {NoStop}%
\bibitem [{\citenamefont {{ATLAS
  Collaboration}}(2013{\natexlab{c}})}]{Aad:2013ucp}%
  \BibitemOpen
  \bibfield  {author} {\bibinfo {author} {\bibnamefont {{ATLAS
  Collaboration}}},\ }\href@noop {} {\bibfield  {journal} {\bibinfo  {journal}
  {Eur. Phys. J. C}\ }\textbf {\bibinfo {volume} {73}},\ \bibinfo {pages}
  {2518} (\bibinfo {year} {2013}{\natexlab{c}})},\ \Eprint
  {http://arxiv.org/abs/1302.4393} {arXiv:1302.4393} \BibitemShut {NoStop}%
\bibitem [{\citenamefont {{ATLAS Collaboration}}(2013{\natexlab{d}})}]{jes11}%
  \BibitemOpen
  \bibfield  {author} {\bibinfo {author} {\bibnamefont {{ATLAS
  Collaboration}}},\ }\href@noop {} {\bibfield  {journal} {\bibinfo  {journal}
  {Eur. Phys. J. C}\ }\textbf {\bibinfo {volume} {73}},\ \bibinfo {pages}
  {2304} (\bibinfo {year} {2013}{\natexlab{d}})},\ \Eprint
  {http://arxiv.org/abs/1112.6426} {arXiv:1112.6426} \BibitemShut {NoStop}%
\bibitem [{\citenamefont {{ATLAS Collaboration}}(2013{\natexlab{e}})}]{jer11}%
  \BibitemOpen
  \bibfield  {author} {\bibinfo {author} {\bibnamefont {{ATLAS
  Collaboration}}},\ }\href@noop {} {\bibfield  {journal} {\bibinfo  {journal}
  {Eur. Phys. J. C}\ }\textbf {\bibinfo {volume} {73}},\ \bibinfo {pages}
  {2306} (\bibinfo {year} {2013}{\natexlab{e}})},\ \Eprint
  {http://arxiv.org/abs/1210.6210} {arXiv:1210.6210} \BibitemShut {NoStop}%
\bibitem [{\citenamefont {{ATLAS
  Collaboration}}(2012{\natexlab{c}})}]{atlasHbb}%
  \BibitemOpen
  \bibfield  {author} {\bibinfo {author} {\bibnamefont {{ATLAS
  Collaboration}}},\ }\href@noop {} {\bibfield  {journal} {\bibinfo  {journal}
  {Phys. Lett. B}\ }\textbf {\bibinfo {volume} {718}},\ \bibinfo {pages} {369}
  (\bibinfo {year} {2012}{\natexlab{c}})},\ \Eprint
  {http://arxiv.org/abs/1207.0210} {arXiv:1207.0210} \BibitemShut {NoStop}%
\bibitem [{\citenamefont {{A.L. Read}}(2002)}]{CLspaper}%
  \BibitemOpen
  \bibfield  {author} {\bibinfo {author} {\bibnamefont {{A.L. Read}}},\
  }\href@noop {} {\bibfield  {journal} {\bibinfo  {journal} {J. Phys. G}\
  }\textbf {\bibinfo {volume} {28}},\ \bibinfo {pages} {2693} (\bibinfo {year}
  {2002})}\BibitemShut {NoStop}%
\bibitem [{\citenamefont {{G. Cowan, K. Cranmer, E. Gross, O.
  Vitells}}(2011)}]{asymptotic}%
  \BibitemOpen
  \bibfield  {author} {\bibinfo {author} {\bibnamefont {{G. Cowan, K. Cranmer,
  E. Gross, O. Vitells}}},\ }\href@noop {} {\bibfield  {journal} {\bibinfo
  {journal} {Eur. Phys. J. C}\ }\textbf {\bibinfo {volume} {71}},\ \bibinfo
  {pages} {1554} (\bibinfo {year} {2011})},\ \Eprint
  {http://arxiv.org/abs/1007.1727} {arXiv:1007.1727} \BibitemShut {NoStop}%
\bibitem [{\citenamefont {{B. Patt and F. Wilczek}}()}]{higgs_portal}%
  \BibitemOpen
  \bibfield  {author} {\bibinfo {author} {\bibnamefont {{B. Patt and F.
  Wilczek}}},\ }\href@noop {} {\ }\Eprint {http://arxiv.org/abs/hep-ph/0605188}
  {arXiv:hep-ph/0605188} \BibitemShut {NoStop}%
\bibitem [{\citenamefont {{P. J. Fox, R. Harnik, J. Kopp, and Y.
  Tsai}}(2012)}]{DM_LHC}%
  \BibitemOpen
  \bibfield  {author} {\bibinfo {author} {\bibnamefont {{P. J. Fox, R. Harnik,
  J. Kopp, and Y. Tsai}}},\ }\href@noop {} {\bibfield  {journal} {\bibinfo
  {journal} {Phys. Rev. D}\ }\textbf {\bibinfo {volume} {85}},\ \bibinfo
  {pages} {056011} (\bibinfo {year} {2012})},\ \Eprint
  {http://arxiv.org/abs/1109.4398} {arXiv:1109.4398} \BibitemShut {NoStop}%
\bibitem [{\citenamefont {{S. Kanemura, S. Matsumoto, T. Nabeshima, and N.
  Okada}}(2010)}]{Kanemura:2010sh}%
  \BibitemOpen
  \bibfield  {author} {\bibinfo {author} {\bibnamefont {{S. Kanemura, S.
  Matsumoto, T. Nabeshima, and N. Okada}}},\ }\href@noop {} {\bibfield
  {journal} {\bibinfo  {journal} {Phys. Rev. D}\ }\textbf {\bibinfo {volume}
  {82}},\ \bibinfo {pages} {055026} (\bibinfo {year} {2010})},\ \Eprint
  {http://arxiv.org/abs/1005.5651} {arXiv:1005.5651} \BibitemShut {NoStop}%
\bibitem [{\citenamefont {{A. Djouadi, O. Lebedev, Y. Mambrini and J.
  Quevillon}}(2012)}]{Djouadi:2011aa}%
  \BibitemOpen
  \bibfield  {author} {\bibinfo {author} {\bibnamefont {{A. Djouadi, O.
  Lebedev, Y. Mambrini and J. Quevillon}}},\ }\href@noop {} {\bibfield
  {journal} {\bibinfo  {journal} {Phys. Lett. B}\ }\textbf {\bibinfo {volume}
  {709}},\ \bibinfo {pages} {65} (\bibinfo {year} {2012})},\ \Eprint
  {http://arxiv.org/abs/1112.3299} {arXiv:1112.3299} \BibitemShut {NoStop}%
\bibitem [{\citenamefont {{R. Bernabei {\it et al.}}}(2008)}]{Bernabei:2008yi}%
  \BibitemOpen
  \bibfield  {author} {\bibinfo {author} {\bibnamefont {{R. Bernabei {\it et
  al.}}}},\ }\href@noop {} {\bibfield  {journal} {\bibinfo  {journal} {Eur.
  Phys. J. C}\ }\textbf {\bibinfo {volume} {56}},\ \bibinfo {pages} {333}
  (\bibinfo {year} {2008})},\ \Eprint {http://arxiv.org/abs/0804.2741}
  {arXiv:0804.2741} \BibitemShut {NoStop}%
\bibitem [{\citenamefont {{J. Angle {\it et al.} (XENON10
  Collaboration)}}(2011)}]{Angle:2011}%
  \BibitemOpen
  \bibfield  {author} {\bibinfo {author} {\bibnamefont {{J. Angle {\it et al.}
  (XENON10 Collaboration)}}},\ }\href@noop {} {\bibfield  {journal} {\bibinfo
  {journal} {Phys. Rev. Lett.}\ }\textbf {\bibinfo {volume} {107}},\ \bibinfo
  {pages} {051301} (\bibinfo {year} {2011})},\ \Eprint
  {http://arxiv.org/abs/1104.3088} {arXiv:1104.3088} \BibitemShut {NoStop}%
\bibitem [{\citenamefont {{E. Aprile {\it et al.} (XENON100
  Collaboration)}}(2012)}]{Aprile:2012}%
  \BibitemOpen
  \bibfield  {author} {\bibinfo {author} {\bibnamefont {{E. Aprile {\it et al.}
  (XENON100 Collaboration)}}},\ }\href@noop {} {\bibfield  {journal} {\bibinfo
  {journal} {Phys. Rev. Lett.}\ }\textbf {\bibinfo {volume} {109}},\ \bibinfo
  {pages} {181301} (\bibinfo {year} {2012})},\ \Eprint
  {http://arxiv.org/abs/1207.5988} {arXiv:1207.5988} \BibitemShut {NoStop}%
\bibitem [{\citenamefont {{G. Angloher {\it et al.}}}(2012)}]{Angloher:2011uu}%
  \BibitemOpen
  \bibfield  {author} {\bibinfo {author} {\bibnamefont {{G. Angloher {\it et
  al.}}}},\ }\href@noop {} {\bibfield  {journal} {\bibinfo  {journal} {Eur.
  Phys. J. C}\ }\textbf {\bibinfo {volume} {72}},\ \bibinfo {pages} {1971}
  (\bibinfo {year} {2012})},\ \Eprint {http://arxiv.org/abs/1109.0702}
  {arXiv:1109.0702} \BibitemShut {NoStop}%
\bibitem [{\citenamefont {{C. E. Aalseth {\it et al.} (CoGeNT
  Collaboration)}}(2011)}]{Aalseth:2011wp}%
  \BibitemOpen
  \bibfield  {author} {\bibinfo {author} {\bibnamefont {{C. E. Aalseth {\it et
  al.} (CoGeNT Collaboration)}}},\ }\href@noop {} {\bibfield  {journal}
  {\bibinfo  {journal} {Phys. Rev. Lett.}\ }\textbf {\bibinfo {volume} {107}},\
  \bibinfo {pages} {141301} (\bibinfo {year} {2011})},\ \Eprint
  {http://arxiv.org/abs/1106.0650} {arXiv:1106.0650} \BibitemShut {NoStop}%
\bibitem [{\citenamefont {{P. J. Fox, J. Kopp, M. Lisanti, and N.
  Weiner}}(2012)}]{Fox:2011px}%
  \BibitemOpen
  \bibfield  {author} {\bibinfo {author} {\bibnamefont {{P. J. Fox, J. Kopp, M.
  Lisanti, and N. Weiner}}},\ }\href@noop {} {\bibfield  {journal} {\bibinfo
  {journal} {Phys. Rev. D}\ }\textbf {\bibinfo {volume} {85}},\ \bibinfo
  {pages} {036008} (\bibinfo {year} {2012})},\ \Eprint
  {http://arxiv.org/abs/1107.0717} {arXiv:1107.0717} \BibitemShut {NoStop}%
\bibitem [{\citenamefont {{R. Agnese {\it et al.} (CDMS
  Collaboration)}}(2013)}]{Agnese:2013}%
  \BibitemOpen
  \bibfield  {author} {\bibinfo {author} {\bibnamefont {{R. Agnese {\it et al.}
  (CDMS Collaboration)}}},\ }\href@noop {} {\bibfield  {journal} {\bibinfo
  {journal} {Phys. Rev. Lett.}\ }\textbf {\bibinfo {volume} {111}},\ \bibinfo
  {pages} {251301} (\bibinfo {year} {2013})},\ \Eprint
  {http://arxiv.org/abs/1304.4279} {arXiv:1304.4279} \BibitemShut {NoStop}%
\bibitem [{\citenamefont {{D. S. Akerib {\it et al.} (LUX
  Collaboration)}}(2014)}]{Akerib:2013}%
  \BibitemOpen
  \bibfield  {author} {\bibinfo {author} {\bibnamefont {{D. S. Akerib {\it et
  al.} (LUX Collaboration)}}},\ }\href@noop {} {\bibfield  {journal} {\bibinfo
  {journal} {Phys. Rev. Lett.}\ }\textbf {\bibinfo {volume} {112}},\ \bibinfo
  {pages} {091303} (\bibinfo {year} {2014})},\ \Eprint
  {http://arxiv.org/abs/1310.8214} {arXiv:1310.8214} \BibitemShut {NoStop}%
\bibitem [{\citenamefont {{ATLAS Collaboration}}(2014)}]{atlas_monoZ}%
  \BibitemOpen
  \bibfield  {author} {\bibinfo {author} {\bibnamefont {{ATLAS
  Collaboration}}},\ }\href@noop {} {\bibfield  {journal} {\bibinfo  {journal}
  {Phys. Rev. D}\ }\textbf {\bibinfo {volume} {90}},\ \bibinfo {pages} {012004}
  (\bibinfo {year} {2014})},\ \Eprint {http://arxiv.org/abs/1404.0051}
  {arXiv:1404.0051} \BibitemShut {NoStop}%
\end{thebibliography}%

\pagebreak 

\onecolumngrid
\begin{flushleft}
{\Large The ATLAS Collaboration}

\bigskip

G.~Aad$^{\rm 48}$,
T.~Abajyan$^{\rm 21}$,
B.~Abbott$^{\rm 112}$,
J.~Abdallah$^{\rm 152}$,
S.~Abdel~Khalek$^{\rm 116}$,
O.~Abdinov$^{\rm 11}$,
R.~Aben$^{\rm 106}$,
B.~Abi$^{\rm 113}$,
M.~Abolins$^{\rm 89}$,
O.S.~AbouZeid$^{\rm 159}$
% ...
\bigskip
\\
% ...
$^{11}$ Institute of Physics, Azerbaijan Academy of Sciences, Baku, Azerbaijan\\
% ...
$^{21}$ Physikalisches Institut, University of Bonn, Bonn, Germany\\
% ...
$^{48}$ Fakult{\"a}t f{\"u}r Mathematik und Physik, Albert-Ludwigs-Universit{\"a}t, Freiburg, Germany\\
% ...
$^{89}$ Department of Physics and Astronomy, Michigan State University, East Lansing MI, United States of America\\
% ...
$^{106}$ Nikhef National Institute for Subatomic Physics and University of Amsterdam, Amsterdam, Netherlands\\
% ...
$^{112}$ Homer L. Dodge Department of Physics and Astronomy, University of Oklahoma, Norman OK, United States of America\\
$^{113}$ Department of Physics, Oklahoma State University, Stillwater OK, United States of America\\
% ...
$^{116}$ LAL, Universit{\'e} Paris-Sud and CNRS/IN2P3, Orsay, France\\
% ...
$^{152}$ Institute of Physics, Academia Sinica, Taipei, Taiwan\\
% ...
$^{159}$ Department of Physics, University of Toronto, Toronto ON, Canada\\
% ...
\end{flushleft}


\end{document}